\documentclass[12pt]{article}

\usepackage{cite,amsmath,epsfig}

\begin{document}
\thispagestyle{empty}

\begin{center}

{\large\sc {\bf SEARCHING FOR HIGGS :  \\[0.4cm]
                FROM LEP TOWARDS LHC{\footnote{Report invited to the special volume 
                                               {\it ``CERN's Accelerators, Experiments and International
                                                    Integration 1959 -- 2009''} of
                                                    Eur. Phys. J. H, edited by  H.~Schopper (CERN Geneva).}} }}
\\[3.5em]
{\normalsize
{\sc
W.-D.~Schlatter$^{1}$
and
P.~M.~Zerwas$^{2}$
}
}

\vspace*{0.75cm}

$^1$ Physics Department, CERN, CH-1211 Geneva, Switzerland

\vspace*{0.4cm}

$^2$ Theory Group, DESY, D--22603 Hamburg, Germany 

\end{center}

\vspace*{2.0cm}

\begin{abstract}

After a brief introduction to the theoretical basis of the Higgs mechanism
for generating the masses of elementary particles, 
the experimental searches for Higgs particles will 
be summarized, from bounds at LEP to inferences for LHC. The report 
will focus on the Standard Model, though some central results on extended 
Higgs systems, as conjectured for example in supersymmetric theories, will 
also be recapitulated. Alternative scenarios based on spontaneous symmetry
breaking by novel strong interactions are adumbrated at the theoretical level.

\end{abstract}

\setcounter{page}{1}
\setcounter{footnote}{0}

\vspace{12mm}

\section{Basics}

\subsection{General introduction}

\noindent
The fundamental laws of nature are formulated, at microscopic distances, 
in the Standard Model [SM] of particle physics \cite{SMGl,SMSa,SMWe,SMGM}. 
The model consists of three elements: \\[-2mm]

-- Leptons and quarks are the basic constituents of matter;  \\[-2mm]

-- The interactions are mediated by gauge fields;            \\[-2mm]  

-- The masses are generated by the Higgs mechanism.          \\[-2mm]

\noindent
Experimental efforts have been devoted over nearly half a century to confront
the theory with the structure of the real world. \\

The particles of the matter sector, fermionic leptons and quarks, arrange themselves
in multiplets associated with underlying symmetry principles. Left-handed
electrons and neutrinos, for example, are paired in doublets, originating
from isospin symmetry, the corresponding right-handed particles however 
remain unpaired in singlets. The particles carry electric 
charges, except for neutrinos, and quarks, in addition, color charges. 
These charges are associated with the symmetry groups SU(3) for color, SU(2) 
for isospin and U(1) for hypercharge, the average electric charge of an 
iso-multiplet, integrated {\it in toto} to the symmetry constellation 
SU(3)$\times$SU(2)$\times$U(1). 
The particles are organized in three families of the same charge structures, 
{\it i.e.} they are isomorphic except for the masses; three families is 
the minimum for accommodating $CP$-violation in the Standard Model. \\[-2mm]

These particles have all been detected experimentally, and their nature,
charges and masses, has been deciphered. Only the properties of the neutrinos
still await clarification. \\[-2mm]

The charges of the matter particles are sources of fields which mediate the
strong, electromagnetic and weak forces among the particles.
The strongly interacting gluons form an octet of color fields.
Mixed combinations of the isospin $W$-fields and the hypercharge $B$ field 
build up the weak and electromagnetic force fields \cite{SMGl}, 
{\it i.e.} the charged and neutral weak fields
$W^\pm,Z^0$ and the electromagnetic photon field $\gamma$. All these fields
are described by gauge theories, a theoretical concept introduced to explain 
the laws of electromagnetism \cite{Weyl}. First developed for neutral force fields,
like the photon, and corresponding to abelian symmetries, they have later been 
generalized to charged force fields, like gluons and weak fields, in non-abelian
theories \cite{YM}. The concept of gauge theories builds the theoretical 
basis of all interactions in the Standard Model. Numerous high-precision 
experiments, {\it notabene} at LEP, have confirmed the predictions derived 
from the gauge symmetries \cite{Erler}, including the self-couplings of the 
color and electroweak gauge fields \cite{nonab}, which are characteristic 
to the non-abelian nature of symmetries. \\[-2mm]
 
The fourth force in nature, gravity, is not isomorphic with the above standard 
forces and of different structure; gravity is attached {\it ad hoc} as a classical 
element to the Standard Model. \\[-2mm]

The color force and the electromagnetic force are of long-range character 
and the associated field quanta, carriers of the forces, are massless,
a straightforward consequence of the gauge symmetries. For quite some
time, the short-range character of the weak force, connected with large
masses of the weak bosons, had been a barrier for describing the weak interactions 
by a non-abelian gauge theory \cite{YM}, a natural symmetry concept for 
forces carrying charges. Masses introduced by hand however destroyed 
the gauge symmetry and thus the very basis of this concept. The problem 
was solved when the concept of gauge symmetry was connected with the concept
of spontaneous symmetry breaking \cite{Nambu,Gold}, in which solutions of the 
field equations have a minor rank of symmetry than the equations themselves. 
The breakthrough for this solution was achieved in 1964 when the Higgs mechanism
was introduced by Englert and Brout \cite{EB}, Higgs \cite{Higgs123}, and Hagen,
Guralnik and Kibble \cite{HaGuKi} to generate masses for vector bosons 
in gauge theories. Even though different techniques were applied, the physical 
key role is played by massless scalar particles which emerge from theories 
with spontaneous symmetry breaking. While the Goldstone theorem \cite{Gold} 
predicts these particles in theories with spontaneously broken global symmetries,
they are absorbed in local gauge theories to build up the longitudinal
field components and to transform the massless gauge bosons to massive states
[with, strictly, the gauge symmetry not broken by the vacuum \cite{gsym}].
The energy transfer to the gauge field connected with the absorption of the 
Goldstone boson can be re-interpreted as generating mass to the gauge
fields. Since the longitudinal field component does not carry spin along
the motion of the field, it can be synthesized by the spinless Goldstone 
boson. \\[-2mm]

This mechanism was adopted for formulating the electroweak sector of the 
Standard Model by Salam \cite{SMSa} and Weinberg \cite{SMWe}. Introducing 
a complex iso-doublet scalar field with four degrees of freedom, as a minimum, 
three Goldstone components of the four scalar field degrees are absorbed to 
provide masses to the $W^\pm,Z$-bosons. At the same time isospin symmetric 
interactions between the fermion fields and the scalar field generate the fermion 
masses. In providing masses to three gauge bosons, one out of the four scalar 
degrees of freedom is left over, manifesting itself as a real neutral scalar 
particle in this basic realization of electroweak symmetry breaking, the 
Higgs boson. \\[-2mm]

Originally introduced into the Standard Model, the Higgs mechanism, as 
a generic element, has been applied subsequently in a large variety 
of theories. They may broadly be divided into two classes. In the first 
class the Higgs field is a fundamental field, eventually up to energies 
close to the Planck scale, the scale of order $10^{+19}$ GeV where
gravity becomes strong and must be intimately connected 
with the particle system. A set of Higgs particles may be incorporated 
in the theories of this class, as required, {\it e.g.}, for supersymmetric 
theories. The second class comprises theories in which novel strong 
interactions, characterized by an energy scale potentially as small as TeV,
trigger the spontaneous symmetry breaking. Basic technicolor theories \cite{TC}, 
which do not include light physical scalar fields, are a characteristic 
paradigm of this class. Intermediate scales and light Higgs bosons identified 
with pseudo-Goldstone bosons are introduced in branches like Little Higgs 
models \cite{LH}. \\

After the early profiling of the Higgs boson in the Standard Model \cite{profile},
a comprehensive picture of the particle was drawn in the {\it ``Higgs Hunter's Guide''}, 
Ref.\cite{HHG}, followed by extensive overviews of electroweak symmetry breaking 
in general and the Higgs mechanism in particular, as presented {\it e.g.} in Ref.\cite{Mon}, 
and by broad discussions of phenomenological aspects{\footnote{We apologize to all authors 
whose important work could not be given the reference proper in this very brief report.}}
in Refs.\cite{Djou} and \cite{PDG}.


\subsection{Higgs in the Standard Model}

\begin{figure}[h] 
\begin{center}
\epsfig{file=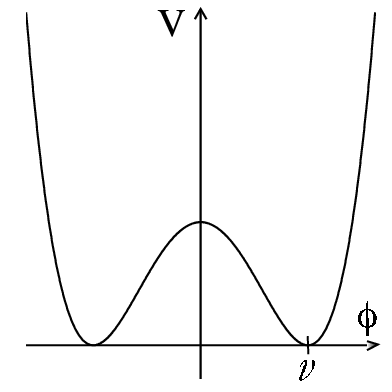,height=4cm}
\end{center}
\caption{\it Characteristic energy potential for scalar fields in theories
             of spontaneously broken symmetry. The ground-state (vacuum) is
             moved from $\phi = 0$ to a non-zero value $v$ at the minimum 
             of the potential.}
\label{fig:mexico}
\end{figure}

\noindent
The Higgs sector of the Standard Model is built up by a scalar iso-doublet
field, including bilinear and quadrilinear self-interactions, in which two
and four scalars, respectively, are coupled with each other. To guarantee
the vacuum to be stable, the coupling $\lambda$ of the quadrilinear interaction 
must be positive. If, on the other hand, the coefficient of the bilinear
mass term is negative, a minimum is induced in the interaction energy and
the ground state shifts from field-strength zero to a non-zero value $v$,
{\it cf.} Fig.~{\ref{fig:mexico}}. 
This ground-state value is related to the Fermi coupling in the weak 
interactions by $v = [1/\sqrt{2} G_F]^{\frac{1}{2}}$, numerically 246 GeV, 
the fundamental electroweak scale. \\

The interaction of the SU(2)$\times$U(1) gauge fields with the scalar 
field $v$ generates gauge-boson masses $M^2 = g^2 v^2 /4$. Similarly
the masses of the fermionic leptons and quarks $f$ are generated by 
gauge-symmetric Yukawa interactions which couple the scalar iso-doublet 
with the fermion iso-doublets and iso-singlets, resulting in fermion 
masses $m_f = g_f v /\sqrt{2}$. \\

In the process of shifting the zero masses of the gauge bosons to non-zero 
values, three Goldstone modes of the scalar field are absorbed by the gauge 
bosons, leaving one physical Higgs state out of the four components of the 
iso-doublet scalar. \\
  

\subsubsection{Higgs mass}

\noindent
The Higgs mass is related to the quadrilinear coupling in the Higgs potential,
$M^2_H = 2 \lambda v^2$. Since the coupling $\lambda$ is not pre-determined, 
the Higgs mass cannot be predicted. In fact, it is the, presently, only unknown 
parameter in the Higgs sector of the Standard Model while all other Higgs couplings
to weak bosons, leptons and quarks can be related to their well measured masses. \\

Constraints, at the theoretical as well as experimental level, 
restrict the value of the Higgs mass quite strongly. A rather general 
bound $M_H <$ 700 GeV follows from the unitarity of the theory \cite{Quigg},
which leads, by requiring the probability for the scattering of particles
to be less than unity, to an upper bound on partial-wave amplitudes in elastic 
$WW$ scattering, depending quadratically on the Higgs mass. However, this 
general bound is reduced considerably if additional theoretical 
assumptions and experimental constraints are exploited. \\ 

\noindent
{\it (a) {\underline{Extrapolation to grand-unification scale}}} \\

\noindent
The quadrilinear Higgs coupling $\lambda$ grows indefinitely with energy,
driven by radiative corrections involving the Higgs self-interactions. To prevent 
the coupling from becoming strong before the scale of the grand unification [GUT]
of the forces of the Standard Model is reached, the value at the electroweak scale 
must be bounded \cite{Maiani}, and the Higgs mass is restricted 
correspondingly, $M^2_H < {8 \pi^2 v^2}/{ 3 \log{(M^2_{GUT}/v^2)}}$.
The underlying assumption of weakly interacting fields in the Standard Model
up to the grand unification scale is supported by the qualitative prediction 
of the electroweak mixing parameter, $\sin^2\theta_W \sim 0.2$, when evolved
from the GUT value $3/8$ down to the value at the experimental electroweak scale
\cite{Quinn}. On the other hand, quantum corrections involving fermionic top 
quarks reduce the quadrilinear Higgs coupling. To prevent the coupling from 
falling below zero, which would render the vacuum unstable, the value at the 
electroweak scale must be bounded from below \cite{Maiani}. {\it In toto}, 
the Higgs mass of the Standard Model is restricted, including theoretical 
uncertainties \cite{Espin}, to the conservative range
\begin{equation}
124\; {\rm GeV} < M_H < 180\; {\rm GeV} 
\end{equation}  
in this scenario. Otherwise new strong interactions would be predicted with 
a characteristic scale between the electroweak scale and the Planck scale. 
Uncertainties of a few GeV in the theoretical estimate of the lower limit
\cite{Espin} must be exhausted if the conflict of a stable vacuum with a Higgs 
mass value of 125 GeV, for example, should be circumvented. \\

\noindent
{\it (b) {\underline{Electroweak radiative corrections}}} \\

\begin{figure}[t]
\centering
\epsfig{file=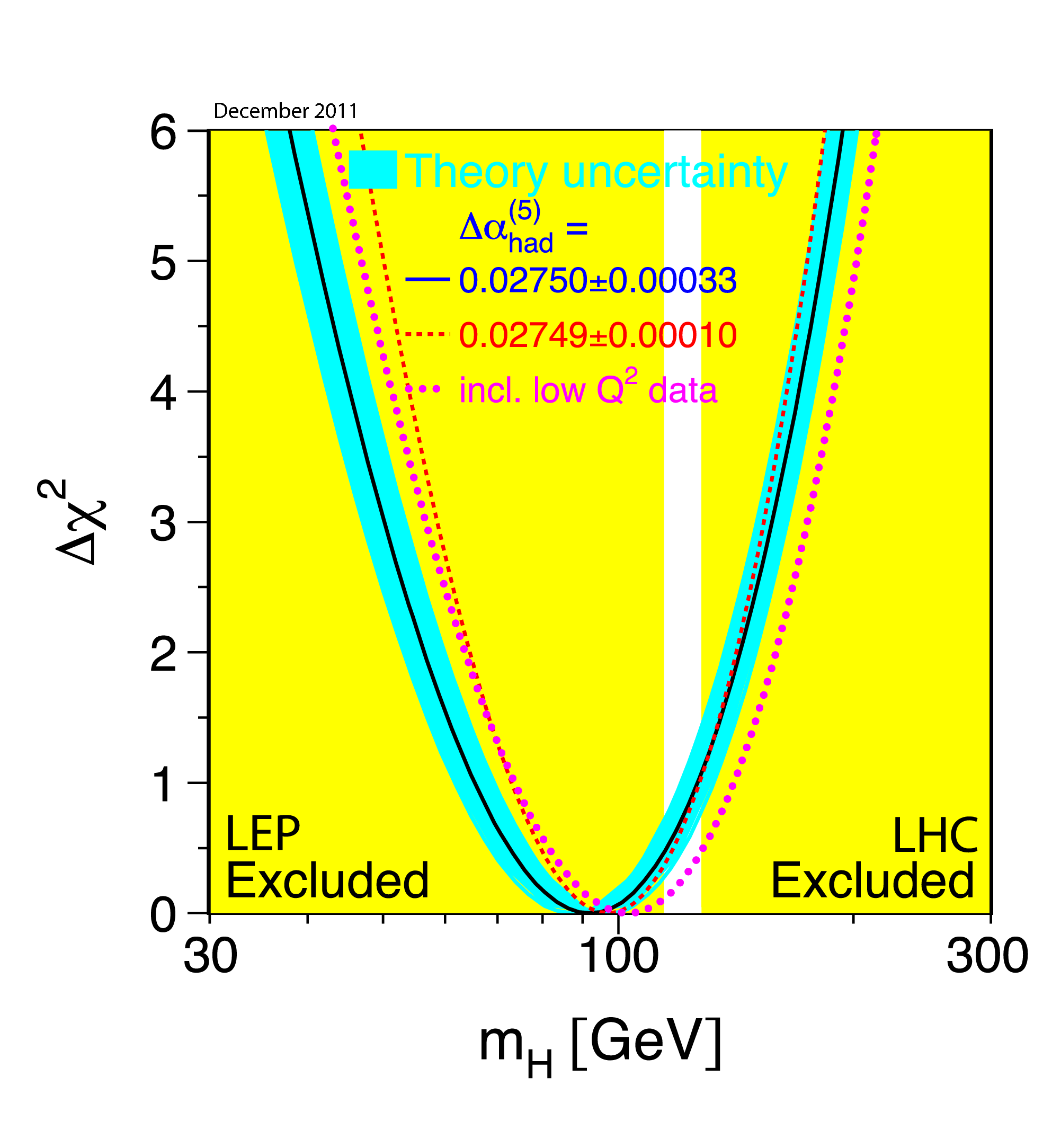,height=9cm}
\caption{\it  Estimate of the Higgs mass in the Standard Model
              from electroweak precision measurements \cite{elwWG}.
			The excluded area labeled LHC has been updated with results
			from Ref.\cite{publicCERNsem}, courtesy M.~Gr\"unewald. Present 
                        search limits for the 
                        Higgs boson from LEP and LHC leave only a small mass gap, the
                        white bar close to the minimum of the $\chi^2$ distribution.}
\label{fig:F_BlueBand}
\end{figure}

\noindent
Support for a light Higgs mass in the Standard Model has also been derived 
from the high-precision measurements of the electroweak observables at LEP1,
the LEP mode which operated around the $Z$ pole, and of observables in
vastly different fields from high-energy neutrino scattering down to polarization
effects in atomic physics. Since the Standard Model is renormalizable 
\cite{HooftVeltman}, {\it i.e.} only a small number of basic parameters, 
masses and couplings, must be introduced in the theory as experimental 
observables, quantum corrections can be predicted to high accuracy. 
The Fermi coupling $G_F$, at the Born level 
$\sim \alpha / \sin^2 2 \theta_{eff}^{lept} M^2_Z$, is affected by quantum 
corrections logarithmically in the Higgs mass \cite{Velt}, $\sim \alpha 
\log {M^2_H}/{M^2_W}$. They add to corrections quadratic in the top-quark mass 
(which had been exploited, in fact, to predict this mass successfully before
top quarks were detected experimentally in $p \bar{p}$ collisions). Since 
all the parameters in $G_F$ were measured at LEP1 very precisely, in particular 
the $Z$-boson mass and the effective mixing parameter $\sin^2\theta_{eff}^{lept}$,
the Higgs mass can be constrained from the experimental data. 
The result of the global analysis \cite{elwWG}, which is displayed in 
Figure {\ref{fig:F_BlueBand}}, can be condensed to the estimates 
\begin{eqnarray}
M_H &=&     92 ^{+34}_{-26} \, {\rm GeV}          \nonumber\\[2mm]
    &\leq& 161 \, {\rm GeV} \; (95\% \, {\rm CL})   
\end{eqnarray}
for expectation value and upper limit of the Higgs mass in the Standard Model, 
to be confronted with direct experimental searches. \\

Thus the GUT-based argument as well as the evaluation of the quantum (radiative)
corrections, largely based on LEP1 measurements, generate a consistent
picture for a light Higgs boson within the Standard Model. \\


\begin{figure}[t]
\centering
\hspace{8mm} \epsfig{file=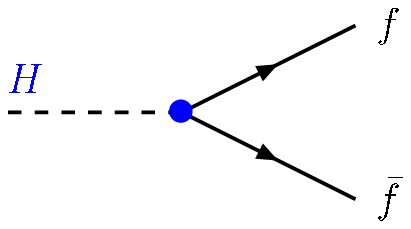,height=1.5cm} \hspace{7mm}  \epsfig{file=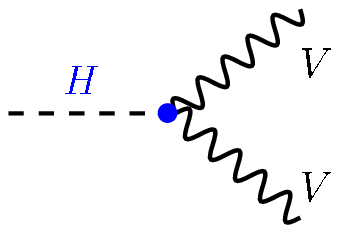,height=1.5cm}  \\[1mm]
\hspace{13mm} \epsfig{file=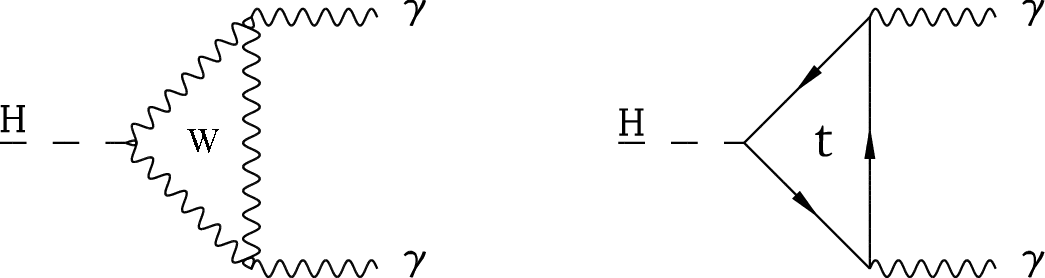,height=1.7cm}                                                  \\[2.5mm]
\epsfig{file=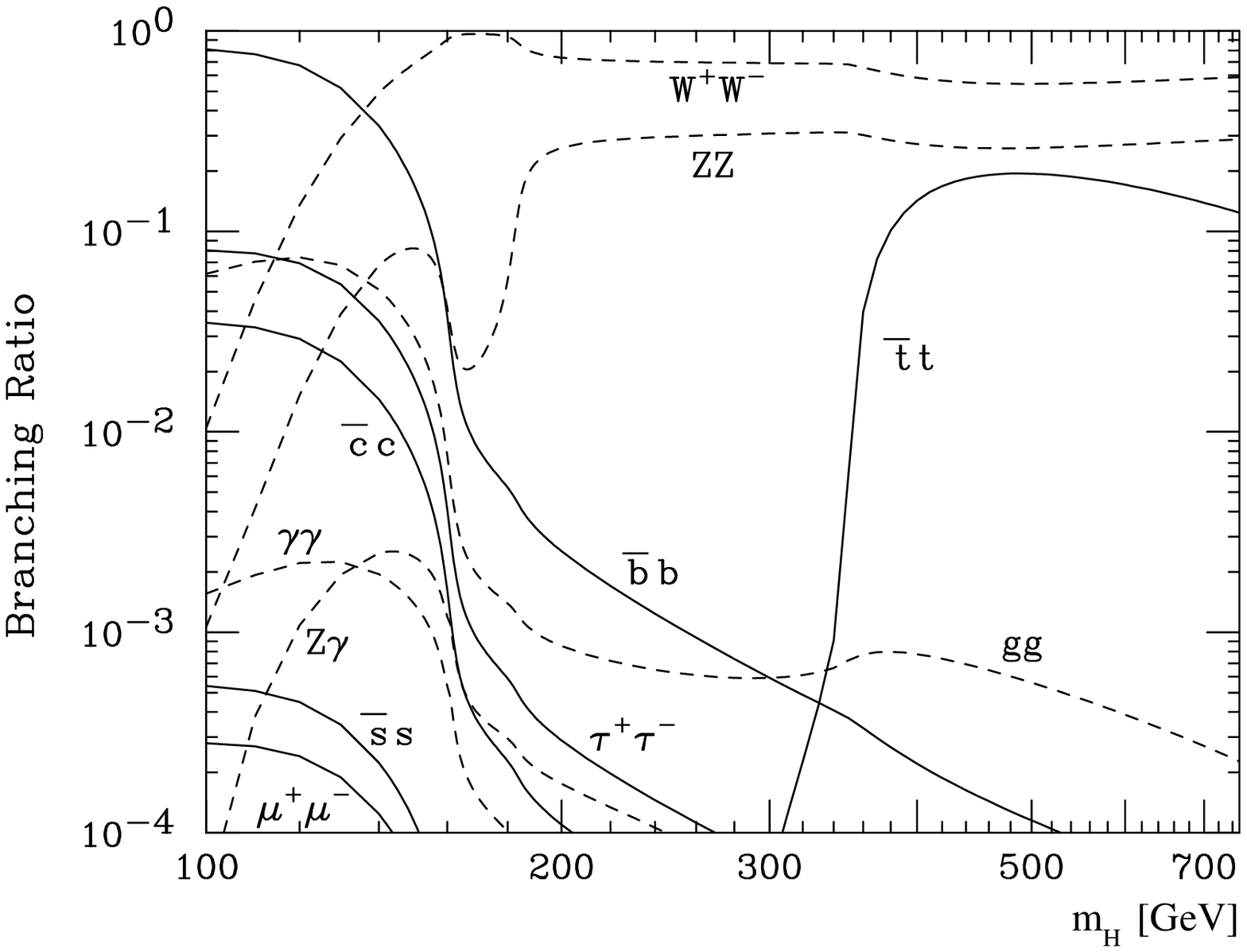,height=7cm}
\caption{\it Branching ratios for Higgs decay modes \cite{Mon}. On top: diagrams
             exemplifying important channels, fermion $f \bar{f}$, vector boson
             $VV = ZZ,WW$ and loop-induced photon $\gamma\gamma$ decay pairs. }
\label{fig:F_HiggsBR}
\end{figure}

\subsubsection{Higgs decay modes and production mechanisms}

\noindent
Conform with the target of the Higgs mechanism, the couplings of the Higgs
boson to pairs of massive gauge bosons and fermions will be of the order 
of their masses: $g_H \sim M_V, m_f$. This rule gives rise to a characteristic 
set of potential production channels in $e^+ e^-$ collisions, and in $p \bar{p}$ 
and $pp$ collisions. Likewise the hierarchy of the decay modes is dictated 
by this mass rule. \\

\noindent
{\it (a) \underline{Decay modes of the Higgs boson}}  \\

\noindent
The decay modes define the mass-dependent signatures in searches for the Higgs 
bosons. The particle with the largest mass kinematically allowed for pair decay 
provides the dominant decay mode, reversed however, by statistics, in the $W,Z$ 
sector and supplemented by photonic (and other) loop decays:
\begin{eqnarray}
&& H \to b \bar{b}, \tau^+ \tau^-; W^+W^-,\, ZZ;\, ...      \\
&& H \to \gamma\gamma                                       \,.
\end{eqnarray}
The branching ratios for the main decay channels are shown in Figure {\ref{fig:F_HiggsBR}}.
Decays to electroweak $W,Z$-bosons are effective already significantly below the pair
thresholds, with one of the bosons being virtual. The Higgs coupling to the massless photons
\cite{profile} is mediated by $W$-boson and $t$-quark loops, as exemplified in the diagrams 
on top of Fig.~{\ref{fig:F_HiggsBR}}. \\

The total width of the Higgs boson remains narrow throughout the low-mass range,
rising from a few MeV near 120 GeV to about 1 GeV at the $ZZ$ threshold. Thus, the width
in this mass range is below the experimental resolution. \\

For Higgs decays to final states without multiple neutrinos and with sufficient background
control, the Higgs boson can fully be reconstructed. Most attractive modes, particularly
in the LHC environment, are the $\gamma\gamma$ decays at low Higgs masses and the 
$ZZ \to \ell\ell\ell\ell$ channel at medium to large masses. These final states generate 
resonance peaks above smooth backgrounds, thus allowing the Higgs reconstruction 
in a model-independent way -- the classical procedure for a particle as fundamental 
as the Higgs boson. \\

\begin{figure}[h]
\centering
\epsfig{file=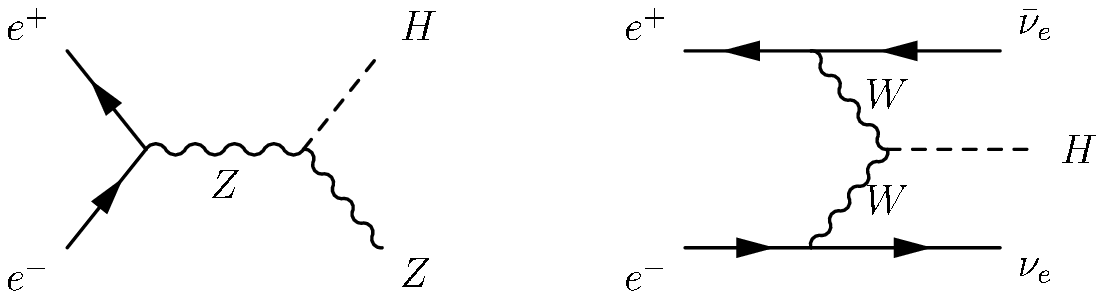,height=2.25cm}    \\[4mm]
\epsfig{file=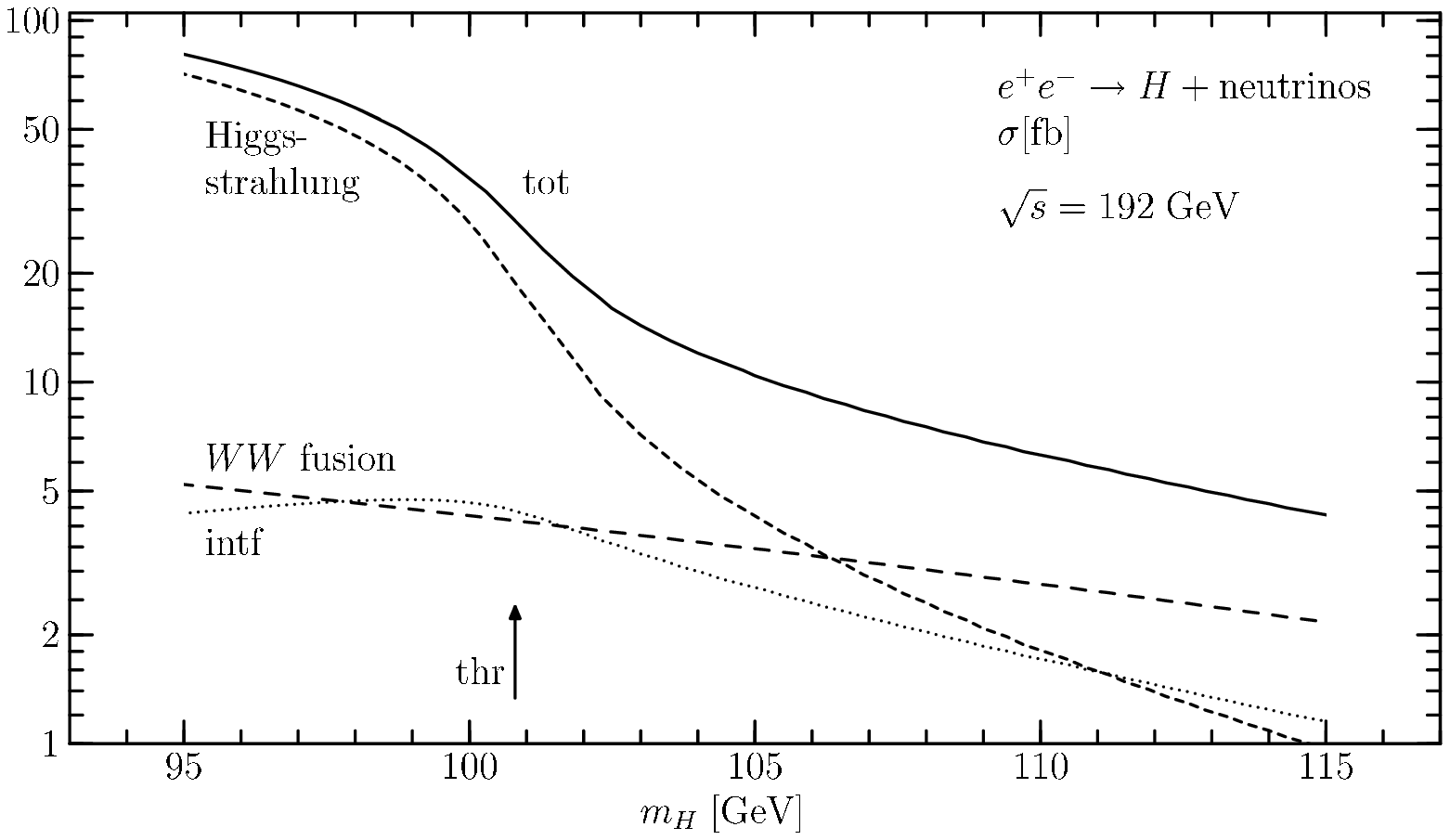,height=7cm}
\caption{\it Higgs production cross section $\sigma[e^+e^- \to H \bar{\nu} \nu]$
             \cite{KKZ}. The total cross section is built up by Higgs-strahlung
             $HZ$ with $Z$ decaying to neutrinos, $WW$ fusion and their interference 
             (intf). Higgs-strahlung falls off rapidly above the threshold (thr) region.
             Other than $\bar{\nu} \nu$ final states in Higgs-strahlung can be
             derived from the short-dashed curve by re-adjusting the $Z$-decay 
             branching ratio properly.
             Diagrams on top describe the Higgs-strahlung and the vector-boson
             fusion mechanism for Higgs production in $e^+e^-$ collisions.}
\label{fig:F_eeH}
\end{figure}

\noindent
{\it (b) \underline{Higgs production in $e^+ e^-$ collisions}} \\

\noindent
Four processes, primarily, have been exploited at LEP to search for Higgs bosons \cite{profile,Bj,
Hstrahl,HWfus,KKZ} without gap from zero-mass up to masses above the $Z$-boson:
\begin{eqnarray}
{\rm LEP1}\; :&&\!\!\!\! {\rm Bjorken}\; {\rm process} {\hspace{3mm}}
                \;\; Z \to H + [Z] \to H + f \bar{f}  \nonumber\\
              &&\!\!\!\! {\rm Radiative}\; {\rm process}      \;\; Z \to H + \gamma                   \\[1.25mm] 
{\rm LEP2}\; :&&\!\!\!\! {\rm Higgs}\!\!-\!\!{\rm strahlung}  \,\;\;\; e^+ e^- \to [Z] \to H + Z      \nonumber\\
              &&\!\!\!\! W{\rm -fusion} {\hspace{11mm}}  
                \,\;\;\; e^+ e^- \to [WW] + \bar{\nu}_e \nu_e \to H + \bar{\nu}_e \nu_e               \,,
\label{eq:Hs}
\end{eqnarray}
the square brackets representing virtual states; additional channels,
like $[ZZ]$ fusion, have been of minor impact. 
While the Higgs boson is coupled to the heavy $Z,W^\pm$-bosons directly,
the coupling to the $Z \gamma$ final state is mediated by virtual $W^\pm$-bosons 
and top quarks. The branching ratios of the Bjorken process 
and the radiative process, and the production cross sections in Higgs-strahlung and 
Higgs $W$-fusion are large enough, {\it cf.} Figure {\ref{fig:F_eeH}}, to cover 
the entire range from zero Higgs mass up to the kinematical limit 
of Higgs-strahlung: $0 < M_H < E^{tot}_{ee} - M_Z$. \\

\noindent
{\it (c) \underline{Higgs production in $pp$ and $p\bar{p}$ collisions}} \\

\begin{figure}[t]
\centering
{\hspace{-6mm}} \epsfig{file=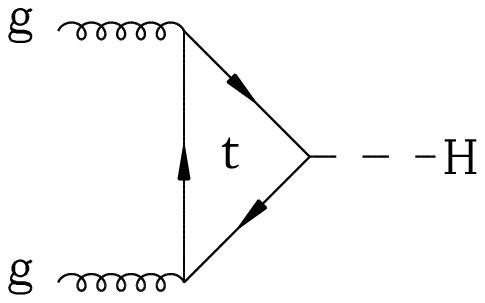,height=2.5cm} \hspace{1cm}
{{\epsfig{file=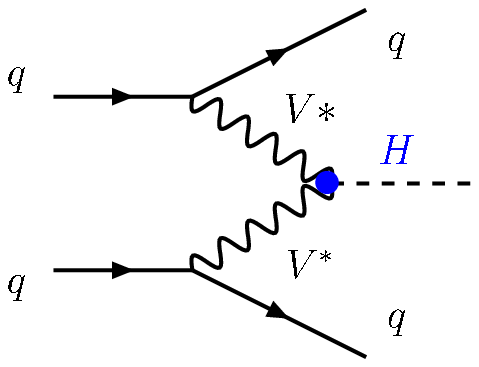,height=2.5cm}}} \\[3mm]
\epsfig{file=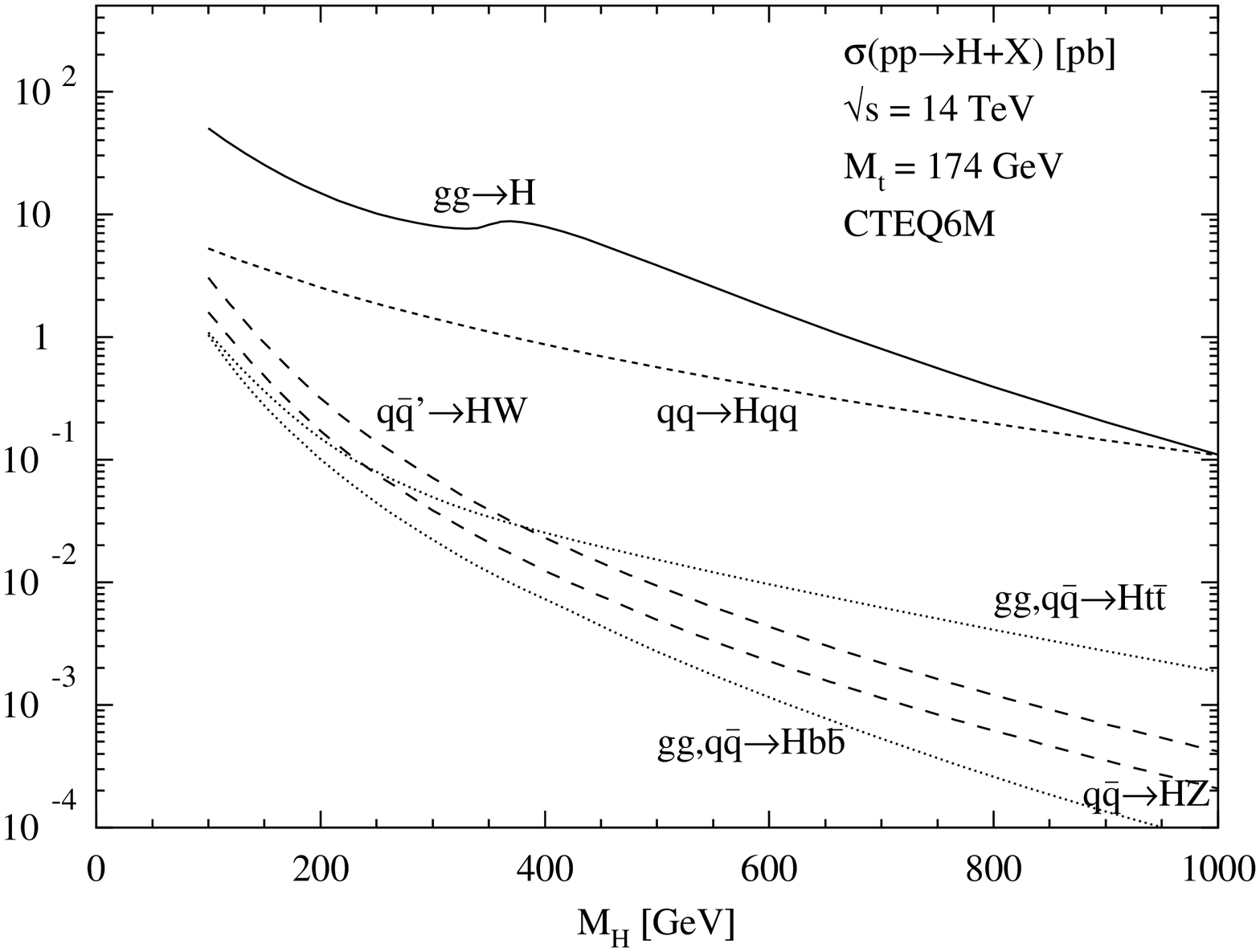,height=8.25cm}
\caption{\it The Higgs production channels at the LHC \cite{Mon}. The value
             of the top mass $M_t$ affects the cross sections for gluon fusion
             and Higgs bremsstrahlung off top-quark pairs; gluon and quark
             densities are adopted from parametrizations of CTEQ6M. On top:
             gluon fusion \cite{ggH0,ggH1,ggH1ew,ggH2,ggH3} and weak boson 
             fusion \cite{VBF}, two central mechanisms for producing Higgs 
             bosons at LHC; the projectiles, gluons and quarks, as fundamental 
             constituents, each carry major fractions of the proton energies.}
\label{fig:F_Higgspp}
\end{figure}

\noindent
The production channels of Higgs bosons in $pp$ collisions at the LHC, and
$p \bar{p}$ collisions at the Tevatron, parallel the LEP2 channels after 
leptons are replaced by quarks, spearheaded however by the gluon-fusion 
mechanism \cite{ggH0,ggH1,ggH1ew,ggH2,ggH3}:
\begin{eqnarray}
{\rm Higgs}\!\!-\!\!{\rm strahlung} &:& q \bar{q} \to [W] \to H + W \quad\; [{\rm and}\; W \to Z]   \nonumber\\
   W{\rm -fusion}                   &:& qq \to [WW] + qq \to H +qq  \quad\; [{\rm and}\; W \to Z]   \nonumber\\
   {\rm top}\;{\rm  bremsstrahlung} &:& gg,qq \to t \bar{t} + H     \nonumber\\ 
   {\rm gluon}\!\!-\!\!{\rm fusion} &:& gg \to H                    \,,    
\end{eqnarray}
The gluons are connected with the Higgs boson by 
a triangular top-quark loop in the gluon-fusion process. The experimental cross sections 
are derived by summing the subprocesses over all quark and gluon parton-densities. 
Controlling of the QCD radiative corrections is particularly important 
for the gluon-fusion process. Next-to-leading order corrections \cite{ggH1} 
raise the cross section by almost a factor 2 compared to the original
leading-order approach \cite{ggH0}; while, similar to the electroweak corrections 
\cite{ggH1ew}, the two-loop order \cite{ggH2} gives rise to a modest additional increment, 
three-loop corrections \cite{ggH3} finally remove residual scale artifacts which are present
when the expansion is truncated at low orders. \\

An overview of the production cross sections for the Higgs bosons in the dominant
channels listed above \cite{ggH0,ggH1,VBF}, and supplemented by the sub-dominant 
channels \cite{HSpp,QQH}, is displayed in Figure~{\ref{fig:F_Higgspp}}. 
For an integrated luminosity $\mathcal{L} = 100$ fb$^{-1}$, the production of 
a large number, more than 1 million, light Higgs bosons is theoretically 
predicted in the Standard Model. \\


\subsection{Extended systems}

\noindent
Spontaneous symmetry breaking has been implemented in a large variety of mechanisms 
for generating the masses of electroweak gauge bosons and fermions. 
Motivated by the successful estimate of the electroweak mixing parameter $\sin^2\theta_W$, 
particular attention has been paid to theories in which the fields remain weakly interacting 
up to energy scales close to the grand unification scale. Supersymmetric theories are the basic 
paradigm of this class of theories. The opposite extreme are theories in which spontaneous 
symmetry breaking is triggered by new strong interactions near the TeV energy scale, 
eventually including composite Higgs bosons. This mechanism has been incorporated first 
in technicolor theories, developed to wider branches later. \\


\subsubsection{Supersymmetry} 

\noindent
Supersymmetry \cite{susy}, a novel symmetry scheme in which fermionic partners 
are associated with the bosonic fields and {\it v.v.}, 
is tightly connected with the Higgs mechanism. 
In theories in which the fields remain weakly interacting up to the GUT scale, 
quantum corrections would generally drive scalar masses \cite{Witten} to values 
close to this scale, $M^2_H \sim \alpha M^2_{GUT}$. In supersymmetric theories
[susy] the fermionic and bosonic corrections however cancel each other, as a consequence 
of the Pauli principle, up to the mass gap between the two types of states, 
{\it i.e.} $M^2_H \sim \alpha (M^2_{susy} - M^2_{SM})$. Neglecting the small
standard masses $M_{SM}$ of the Standard Model, for low Higgs masses near $M_Z$ 
the masses of supersymmetric partners should therefore not exceed the scale $M_{susy} < 
\mathcal{O}$(1~TeV). \\

To guarantee the theory to be supersymmetric, the Higgs iso-doublet of
the Standard Model must at least be doubled. The doubling expands the number 
of real Higgs states to five: three neutral particles and one pair of charged 
particles, $\{h^0,H^0,A^0,H^\pm\}$, {\it cf.} Ref.~\cite{HaGu}. Since the quadrilinear 
coupling in the Higgs potential of the minimal theory is set by the square of the gauge 
couplings, the mass of the lightest neutral particle $h^0$, after including 
radiative corrections of the order of $G_F m_t^4$, is predicted \cite{susyho} 
to be small, about
\begin{equation}
M_{h^0} < 135\,{\rm GeV} \,,
\end{equation} 
while the masses of the remaining particles $\{H^0,A^0,H^\pm\}$ may have values
anywhere between the electroweak scale and $\mathcal{O}$(TeV).   \\

The production cross sections of the lightest Higgs boson $h^0$ for
Higgs-strahlung and $W-$fusion in $e^+e^-$
collisions are reduced $\sim \sin^2(\alpha - \beta)$ in relation to the Standard Model 
by the mixing of Higgs states among themselves (angle $\alpha$) and with Goldstone states 
(angle $\beta$) \cite{HaGu}. The suppression is balanced partly, however, by the additional 
pair-production channel $e^+e^- \to h^0 A^0$ of size $\sim \cos^2(\alpha - \beta)$. Similarly 
for LHC production processes. For the mixing parameter $\tan\beta$ sufficiently
large, Higgs bremsstrahlung off $b-$quarks, produced pairwise at LEP and LHC, 
provides another potentially copious source of Higgs bosons. \\

While supersymmetry solves the problem to keep Higgs masses low in weakly interacting 
theories in a natural way, large supersymmetry scales, as signaled by rising bounds 
on masses of supersymmetric particles, indicate however another, yet much less severe 
problem. In the minimal supersymmetric standard model, the supersymmetric Higgs mass 
parameter $\mu$ and the mass parameters $M_k$ breaking supersymmetry in the 
scalar sector, both of TeV size, must nearly cancel each other, though being 
of unrelated origin, to generate the small electroweak scale, $\frac{1}{2} M^2_Z 
\simeq  c_k M_k^2 - \mu^2$. Extending the minimal supersymmetric model 
by additional Higgs fields, iso-scalars for example, can ease the problem. 
Thus, the Higgs sector in supersymmetry may have a rather complex structure. \\


\subsubsection{Strong electroweak symmetry breaking}

\noindent
Spontaneous symmetry breaking is a well established concept in strong interactions,
giving rise to zero-mass pions if small quark masses are neglected. 
Introducing new strong technicolor interactions, in parallel to QCD but at a high scale 
$\Lambda_{TC} ~\sim$ TeV \cite{TC}, will lead, in analogy, to Goldstone bosons, massless bound states 
of new fermions,  which are absorbed by electroweak gauge bosons to generate non-zero masses, 
$M^2 = g^2 f^2 /4$ with $f = v$. This mechanism does not incorporate any light Higgs particles 
{\it sui generis}, but physical scalar masses are in general of size TeV. Interactions between 
$W,Z$ bosons become strong at TeV energies, affecting the predictions for quasi-elastic $WW$ 
scattering amplitudes \cite{Bagg}. Fermion masses demand the systematic extension of the theory. 
Dynamical solutions, which deviate, as suggested in walking technicolor, from the standard QCD path, 
are required however to reconcile the theory with the observed suppression of flavor-changing processes 
and the precision measurements at LEP. \\

Two scales characterize technicolor theories, the technicolor scale $\Lambda_{TC}$
and the scale characterizing spontaneous symmetry breaking which coincides with the electroweak 
scale $v$. This concept can accommodate also light Higgs bosons if the theories are extended in a form
as realized in Little Higgs Models \cite{LH}. Introducing new strong interactions at a scale $\Lambda_\ast$ 
of 10 TeV or beyond, the spontaneous breaking of a global symmetries associated with the new interactions
gives rise to a large set of Goldstone bosons in addition to the standard iso-doublets. 
Gauge symmetries of the theory extended beyond the standard gauge group are broken at the 
same time, leading to new gauge bosons at the TeV scale. Quantum corrections endow most
of the Goldstone bosons with masses of order TeV, while the standard iso-doublet 
Higgs boson will acquire mass only by multiple quantum corrections, generating the small 
standard electroweak scale $v$. Thus, in addition to the spectrum of the Standard Model 
including only one light Higgs boson, models of this type generate extended spectra of new gauge bosons 
and Higgs bosons, as well as fermions, at the TeV scale, which can be searched for at the LHC
\cite{Han}. \\

Aspects of technicolor models can be connected with extra space dimensions \cite{Xtra,D5}. 
The fifth components of the gauge fields in theories, in which 4-dimensional space-time is extended 
by a new space dimension, can be identified with the scalar fields. Proper boundary 
conditions on the gauge fields in the extra dimension generate electroweak symmetry breaking.
The experimental observation of Kaluza-Klein states in the TeV energy range at LHC, coming
with the compactification of the five dimensions to the standard four space-time dimensions, 
would open the gate to this scenario of electroweak symmetry breaking. \\

\vspace{6mm}

\section{Experimental Results}
\newcommand{\mHrec}{ $m^{\rm rec}_{H}$ \ }
\newcommand{\epem}{$e^+e^-$}
\newcommand{\Hbb}{$H \to b\bar{b}$}

The Higgs bosons in the Standard Model and extended theories have been searched for at
the high-energy $e^+ e^-$ collider LEP and the hadron $p \bar{p}$ collider Tevatron; 
the search is presently continuing at the $pp$ collider LHC. The experiments have
been built on a solid theoretical basis, initiated in early pioneering work and
elaborated to comprehensive pre-studies of the experimental programs developing 
at both types of colliders, see {\it e.g.} Refs.~\cite{Cee} and \cite{Cpp}. \\ 

The search for the neutral Higgs boson in the Standard Model was an integral part of the physics programme of the four LEP collaborations, ALEPH, DELPHI, L3 and OPAL, from the running of LEP at the $Z$ pole (LEP1) up to the highest centre-of-mass energies, in particular from 206 GeV to 209 GeV (LEP2). \\

Data collected by the four LEP collaborations prior to the year 2000 gave no direct
indication of the production of the Standard Model Higgs boson \cite{exp_prel4} 
and excluded masses from zero  up to a lower bound of 107.9 GeV (at 95\% confidence level). During the last year of
the LEP programme, substantial data samples were collected at centre-of-mass
energies exceeding 206 GeV, extending the search sensitivity to Higgs boson masses of about
115 GeV through the Higgs-strahlung process $e^+e^- \to HZ$ (see Eq.(\ref{eq:Hs}))  and providing one of the most stringent limits for the Higgs mass. \\

After the final results from the four collaborations were published individually \cite{exp_9-12}, a LEP-wide combination 
was performed in order to increase the overall sensitivity of the searches for a possible Higgs boson signal. The result is published in \cite{exp_ADLO2003} and is summarized here
{\footnote{For this combination a special working group was created by the four Collaborations \cite{LEP-Higgs-WG}.}}.
The data were collected for centre-of-mass energies from 189 GeV to 209 GeV corresponding to a total integrated luminosity of  $\approx 2.5$ fb$^{-1}$. \\

\subsection{Standard-Model Higgs boson search}

\subsubsection{Search procedures}
At LEP energies, the Standard-Model Higgs boson would be produced mainly in association with
the $Z$ boson through the Higgs-strahlung process $e^+e^- \to HZ$. 
This process is supplemented by a small contribution from the $W$-fusion, which produces a Higgs boson and 
a pair of neutrinos in the final state (see Eq.(\ref{eq:Hs}) and Figure \ref{fig:F_eeH}).  
As illustrated in Figure \ref{fig:F_HiggsBR} for masses below 115 GeV the Higgs boson would decay mainly into $b \overline{b}$ 
quark pairs with a branching ratio of 74\% 
while decays to $\tau^+ \tau^-, W^+W^-, \ gg \ \approx 7\%$ each, and $c\bar{c}$ $\approx 4\%$ 
constitute the rest of the decay width. 
The dominant final-state topologies are $b$-jets, quark jets and lepton pairs as produced in
 the following channels: 
\begin{itemize}
	\item four jet final state (\Hbb)($Z \to q\bar{q}$), 
	\item leptonic final states (\Hbb)($Z \to \ell^+\ell^-$), where $\ell$ denotes charged 
		leptons $e$, $\mu$, $\tau$
	  \footnote{Tau pairs can also contribute through the 
	   $(H \to  \tau^+ \tau^-)(Z \to q{\bar{q}})$ final state.},
	\item  missing energy final state (\Hbb)(${Z \to \nu\overline{\nu}}$).
\end{itemize}

\begin{figure}[htp]
	\centering
	\epsfig{file=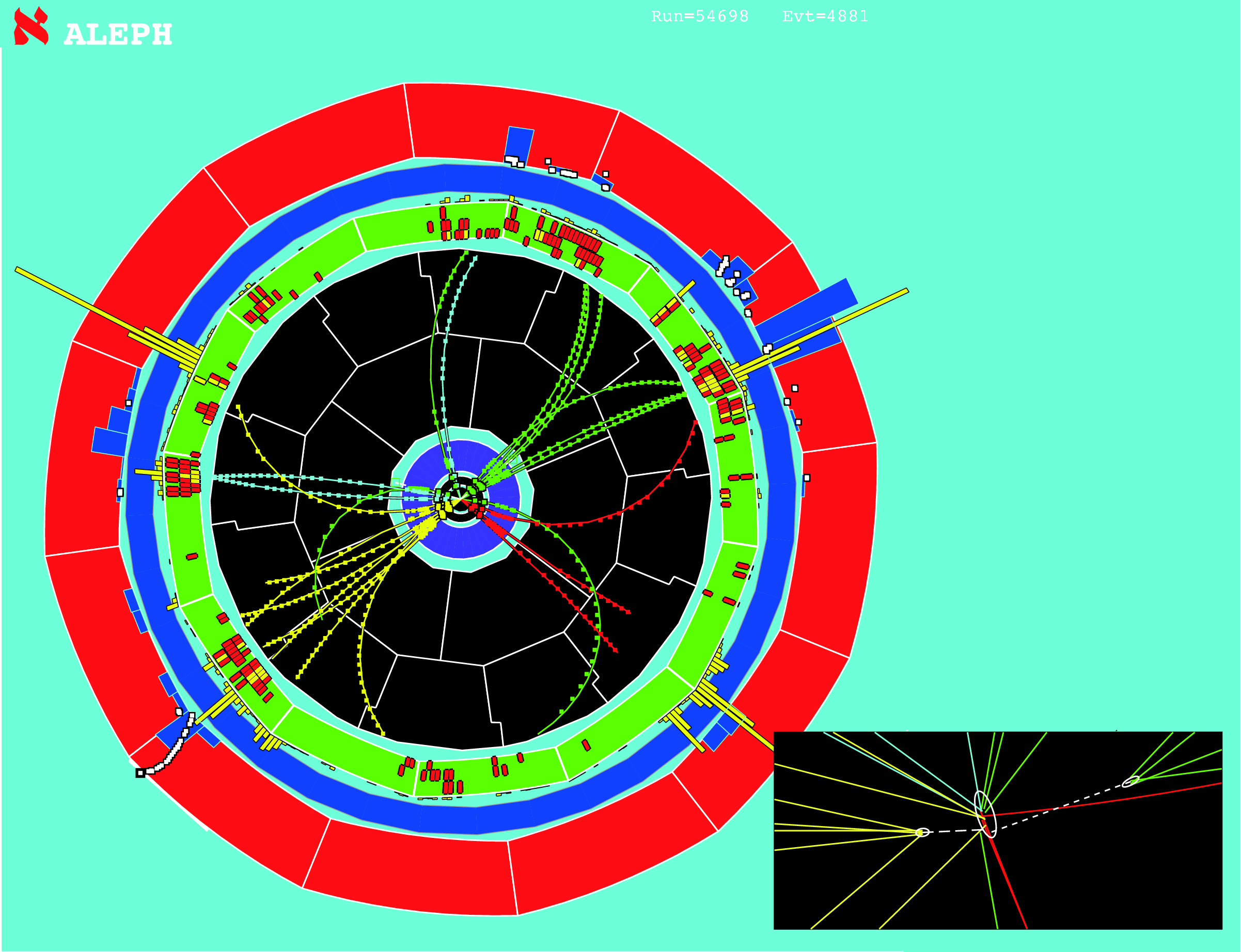,height=7cm} 
	\caption{\it Four-jet event, shown in the view transverse to the beam. The insert shows a closeup of the charged particles at the interaction region. Two secondary vertices are clearly reconstructed, consistent with two decaying b-quark particles.}
	\label{fig:ALEPHevt}
\end{figure}

Background from two-photon processes and from the radiative return to the $Z$ boson, \epem $\to Z\gamma(\gamma)$  fermion pairs 
and $WW$ or $ZZ$ production is effectively reduced by selective cuts and the application of multivariate techniques 
such as likelihood analyses and neural networks. The identification of $b$-quarks in the decay of the Higgs boson,
by secondary vertices for instance, 
is essential in the discrimination between signal and background.
The details of these analyses by the four experiments are given in \cite{exp_9-12}. In Figure  \ref{fig:ALEPHevt} an example of a candidate event from one of the LEP experiments in the four-jet channel (a $b\bar{b}$ pair together with a $q\bar{q}$ pair) is shown. \\

The data from the four experiments are combined using subsets of specific final-state topologies or
of data sets collected at different centre-of-mass energies. 
Essentially, two variables are used to statistically discriminate between signal and background events.
\\

\begin{figure}[htb]
	\centering
	\epsfig{file=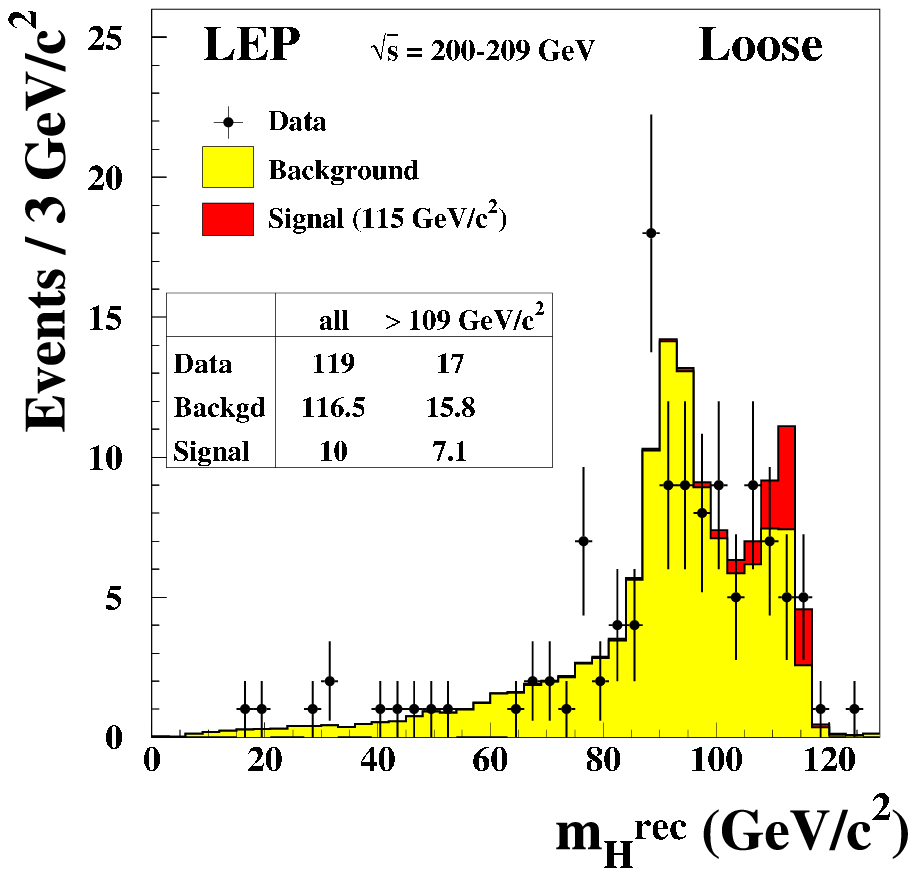,scale=.8} 
		\hfill   \epsfig{file=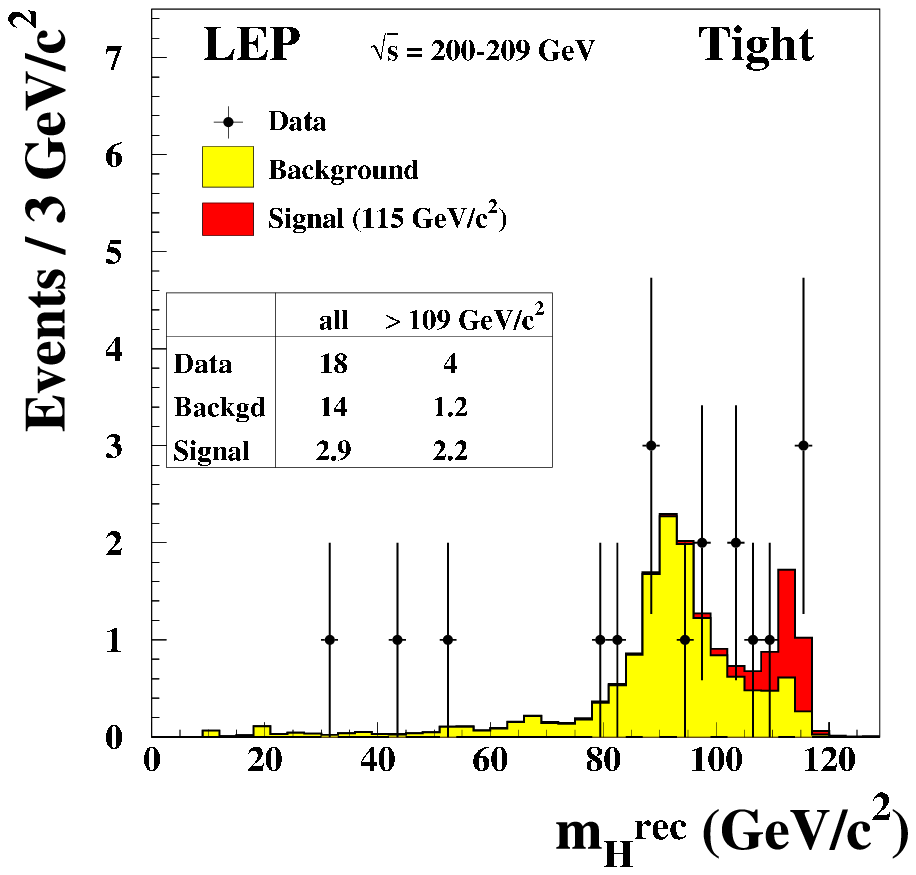,scale=.8}
	\caption{\it Distributions of the reconstructed Higgs boson mass
	 	for two selections with different signal purities. 
		Superimposed to the data (dots) are the Monte Carlo predictions (histograms),
		lightly shaded for the background, heavily shaded for a Higgs boson of mass 115 GeV.
		In the loose and tight selections the cuts are adjusted in such a way as to
		 obtain, for a Higgs boson of mass 115 GeV, approximately 0.5 or 2 times more expected signal
		 than background events when integrated over the region \mHrec $\geq$ 109 GeV \cite{exp_ADLO2003}.
		}
	\label{exp_fig5}
\end{figure}

The most discriminating variable is the reconstructed Higgs boson mass $m_H^{\rm rec}$, in the distribution of which 
the signal events would accumulate around the real Higgs boson mass. 
In Figure \ref{exp_fig5} the observed distributions together with the expectations are shown for illustration at two 
levels of signal purity. Reasonable agreement of the data with the expected background is observed. The enhancement  
in the vicinity of the $Z$ boson mass is from the $ee \to ZZ$ background process.
However, for Higgs boson production close to the kinematic threshold, only few events are expected and therefore a 
higher level of discrimination is needed. \\

The second variable used in the analysis combines many discriminating event features which allow to distinguish on a 
statistical basis between signal and background events. The most important of these variables describes the 
identification of a $b$-quark which appears in all Higgs search channels (see above).  
This information is combined with other event features like multivariate 
tests, likelihood functions or neural networks, leading together to the second discriminating variable that is used in the 
combined analysis. 

\subsubsection{Hypothesis testing}
The combined LEP dataset is tested for compatibility with two hypotheses:  
the background-only scenario and the signal-plus-background scenario. 
The two discriminating variables described in the previous section are used to calculate 
the likelihoods ratio
\begin{equation}
 Q = L_{s+b}/L_b 	
\end{equation}
which is defined with the likelihood functions $L_{s+b}$  and $L_b$ given by the probability density function of the two hypotheses evaluated at the data points. 
For convenience, the logarithmic form $-2\, \ln\, Q$ is used as the test statistic since this quantity 
is approximately equal (exactly equal in the limit of high statistics) to the difference in $\chi^2$ when the 
observed distribution in the two variables is compared to that expected on the basis of simulations for the two hypotheses.
\\

\begin{figure}[h]
	\centering
	\epsfig{file=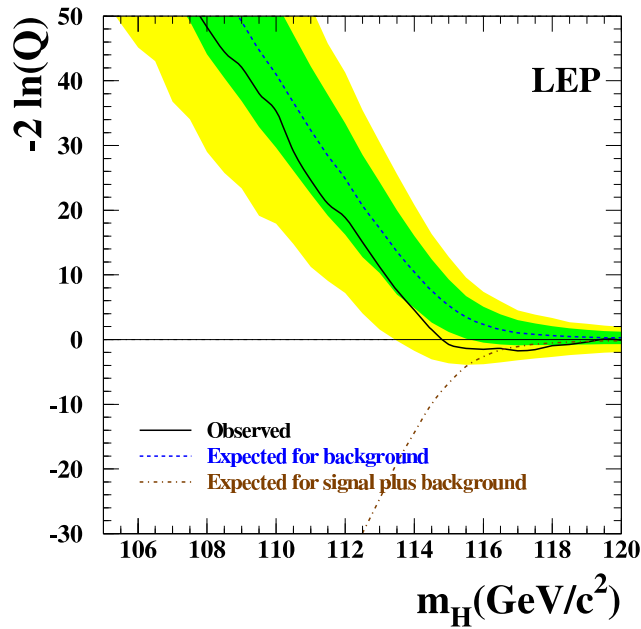,scale=.6}
	\caption{\it The log-likelihood estimator $-2\, \ln\, Q$ as a function of the Higgs boson mass. 
		The full curve represents the
		observation; the dashed curve shows the median background expectation; the dark and light shaded
		bands represent the 68\% and 95\% probability bands about the median background expectation. The
		dash-dotted curve indicates the position of the minimum of the median expectation for the 
		signal-plus-background hypothesis \cite{exp_ADLO2003}.
		}
	\label{exp_fig1}
\end{figure}

In Figure \ref{exp_fig1} the likelihood test $-2\, \ln\, Q$ is shown as a function of the hypothesized Higgs boson mass $M_H$.

Values of $-2\, \ln\, Q$ larger than zero correspond to a likelihood ratio less than one, {\it i.e.} 
the likelihood for the signal-plus-backgound hypothesis is smaller than the one for background only.
The small negative values of $-2\, \ln\, Q$ above 114~GeV
indicate that the hypothesis including a Standard Model Higgs boson of such a mass is slightly more
favoured than the background hypothesis. 
Also, the median expectation for the signal-plus-background hypothesis converges to the observation in this mass range. \\

The compatibility of a data set with a hypothesis is quantified by the
confidence level CL. Here, CL is the probability of obtaining in simulated
measurements a likelihood ratio, as defined above, smaller than the one
observed with the data. If CL$_b$ is the confidence level for the
compatibility of the data in a Higgs search with the background-only
hypothesis, the $p$-value \cite{Cowan}, $p = 1 - {\rm CL}_b$, is,
correspondingly, the probability to obtain a configuration of events
which is less background-like than the one observed. A signal would produce
an excess relative to the background, which would appear as a dip
in $1 - {\rm CL}_b$. 
\\

\begin{figure}[h]
	\centering
	\epsfig{file=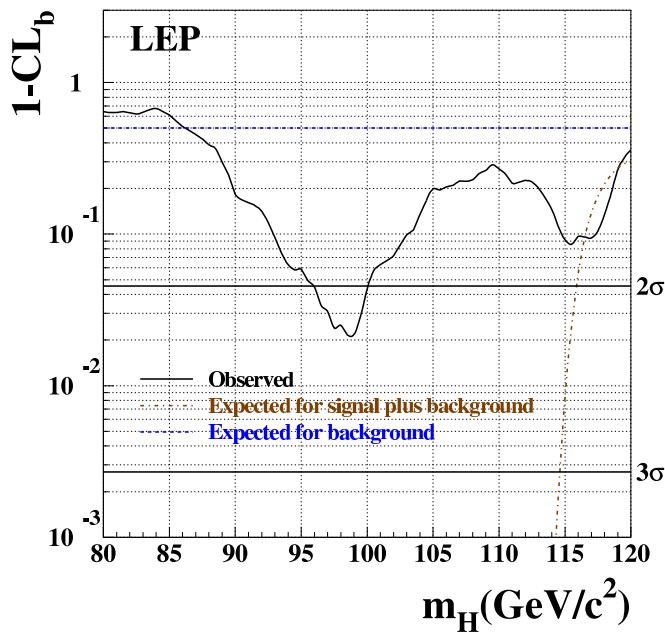,scale=.55}
	\caption{\it The background confidence $1-{\rm CL}_b$ as a function of the test mass $M_H$. Full curve:
		observation; dashed curve: expected background confidence; dash-dotted line: the position of the
		minimum of the median expectation of $1-{\rm CL}_b$ for the signal-plus-background hypothesis, when
	 the signal mass indicated on the abscissa is tested. The horizontal solid lines indicate the levels
		for $2\sigma$ and $3\sigma$ deviations from the background hypothesis \cite{exp_ADLO2003}.}
	\label{exp_fig7}
\end{figure}

Figure \ref{exp_fig7} shows the background confidence $1-{\rm CL}_b$ for $M_H$ in the range from 80
to 120~GeV. The dip in the region $M_H$ $\approx$ 98~GeV corresponding to 2.3
standard deviations,  cannot be interpreted as coming from the Standard Model Higgs boson. 
The number of signal events which would produce such a deviation from the background expectation 
is about an order of magnitude smaller than the number expected for a Standard Model Higgs boson. 
In the region of $M_H$ $\approx$ 115~GeV the dip is compatible with the Standard Model Higgs boson hypothesis 
but its significance is only of 1.7 standard deviations. 
\\ 

The dash-dotted line in Figure \ref{exp_fig7} shows the position of the minimum of the median expected
$1-{\rm CL}_b$ for the signal-plus-background hypothesis. This line indicates the depth of the dip that would result from a Standard Model Higgs boson of mass $M_H$. \\

\subsubsection{Higgs mass limit}

\begin{figure}[h]
	\centering
	\epsfig{file=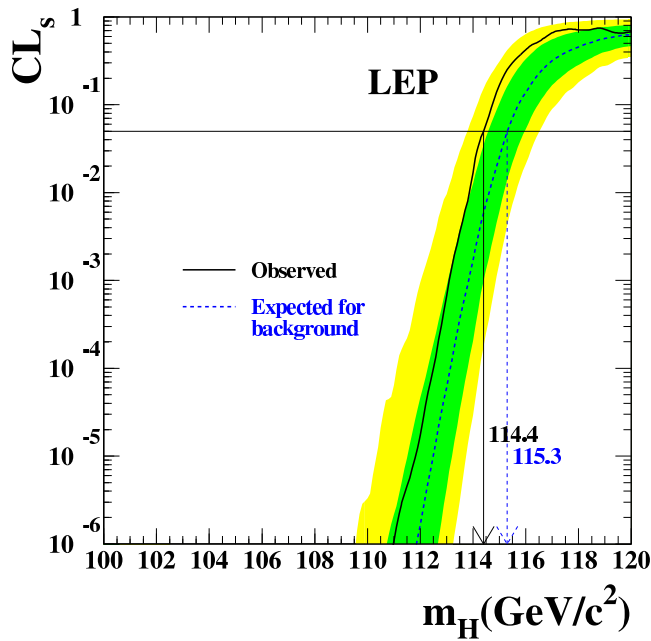,scale=.55}
	\caption{\it The ratio ${\rm CL}_s = {\rm CL}_{s+b}/{\rm CL}_b$ for the signal-plus-background hypothesis. 
                Solid line: observation;
		dashed line: median background expectation. The dark and light shaded bands around
		the median expected line correspond to the 68\% and 95\% probability bands. The intersection of the
		horizontal line for CL$_s$ = 0.05 with the observed curve is used to define the 95\% 
		confidence level lower bound on the mass of the Standard Model Higgs boson \cite{exp_ADLO2003}.}
	\label{exp_fig9}
\end{figure}

Similarly, one can also quantify the compatibility of the observation with the signal-plus-background hypothesis, ${\rm CL_{s+b}}$.
The confidence level ratio ${\rm CL}_s = {\rm CL}_{s+b}/{\rm CL}_b$ is a function of the test mass and is used to derive a lower bound 
on the Standard Model Higgs boson mass (Figure \ref{exp_fig9}). The lowest mass giving ${\rm CL}_s$ = 0.05 is taken 
as the lower bound at the 95\% confidence level. 
The observed 95\% CL lower bound on the mass of the Standard Model Higgs boson obtained from
LEP data is 114.4~GeV while the expected 95\% CL is 115.3~GeV. \\

\subsubsection{Towards the LHC}

After the closure of the LEP collider, the search for the Higgs boson was continued at the Fermilab Tevatron $p\bar p$ collider. 
The cross section for the production of the Standard Model Higgs boson in $p\bar p$ collisions is close to 1 pb for low masses. 
The two experiments, CDF and D0, have not observed any excess and exclude the region of 156~GeV $\leq M_H \leq$ 177~GeV at 95\% CL 
\cite{exp_CDF-D0-comb}. \\

Since 2010, the CERN $pp$ collider LHC is in operation, in the first phase with an energy of 7 TeV, to be extended later
to 14 TeV. Hence the focus of the Higgs boson search has moved to the two large detectors ATLAS and CMS 
\cite{exp_ATLAS2011, exp_CMS2010}. They have been optimized for the Higgs boson search in a mass range from the LEP limit 
of 114.4~GeV up to $\approx$ 700~GeV. In $pp$ collisions the Standard Model Higgs boson production is dominated by gluon-gluon fusion  
(Figure \ref{fig:F_Higgspp}) with a large cross section of tens of pb. \\[-1mm]

The dominant Higgs decay mode below $\approx$ 140 GeV is $H \to b\bar{b}$, 
and above $\approx$ 140~GeV it is $H \to WW \ ({\rm and} \ ZZ)$.  
However, for a mass below 130~GeV the most sensitive channel in terms of signal to background discrimination  is the 
$H \to \gamma\gamma$ final state despite its low branching ratio of $\approx 2 \times 10^{-3}$.  Above 130~GeV 
the $H \to WW$ and $H \to ZZ$  final states dominate.  At high masses the $H \to ZZ$ mode with each $Z$ decaying 
to an $e^+e^-$ or $\mu^+\mu^-$ pair is the cleanest channel. Irreducible backgrounds in the most sensitive 
final states $WW$, $ZZ$ and $\gamma\gamma$ are from $q\bar q$ annihilation processes. \\[-1mm]

In December 2011 ATLAS and CMS reported in a special seminar at CERN\cite{publicCERNsem} the status on the Higgs boson search based on data collected by each experiment corresponding to an integrated luminosity of about 5~fb$^{-1}$. The main result is that the Standard Model Higgs boson, if it exists, is most likely to have a mass constrained to the range 
$$ 115.5 - 131~{\rm GeV ~(ATLAS)}$$ 
$$ 115 - 127~{\rm GeV ~(CMS)}.  $$
Intriguing hints have been seen by both experiments in the mass range of 124 - 126~GeV, in particular 
in the $H \to \gamma\gamma$ channel {\footnote{The mass of such a particle would fit to expectations within 
the Standard Model \cite{SM125a} though positive identification with the SM Higgs boson \cite{SM125b} 
would be a long-time process; 
analogously for supersymmetric interpretations, see {\it e.g.} Refs.~\cite{SUSY125}.}}. However, the excess in 
both experiments is  not strong enough to claim a discovery. Hence, the hunt for the Higgs boson will continue 
at the LHC during 2012. \\[-1mm]

With high luminosity at the LHC, the experiments have the potential to fully cover the  mass range 
for the Standard-Model boson search. \\[-1mm]

\subsection{Supersymmetry}

In supersymmetric theories, like the Minimal Supersymmetric Standard Model MSSM, 
the lightest of the neutral Higgs bosons, $h^0$, is
naturally predicted to have a mass less than $\approx$ 135~GeV (see chapter 1.3.1). 
This prediction provided
a strong motivation for the searches at LEP energies. The masses of the other $CP$-even
and odd neutral and charged bosons Higgs bosons $H^0, A^0, H^\pm$ in the MSSM (and
supersymmetric theories in general) may be as large as $\mathcal{O}$(1 TeV), the
typical mass scale of supersymmetric theories. \\[-1mm]

\begin{figure}[t]
        \centering
                \epsfig{scale=.5,file=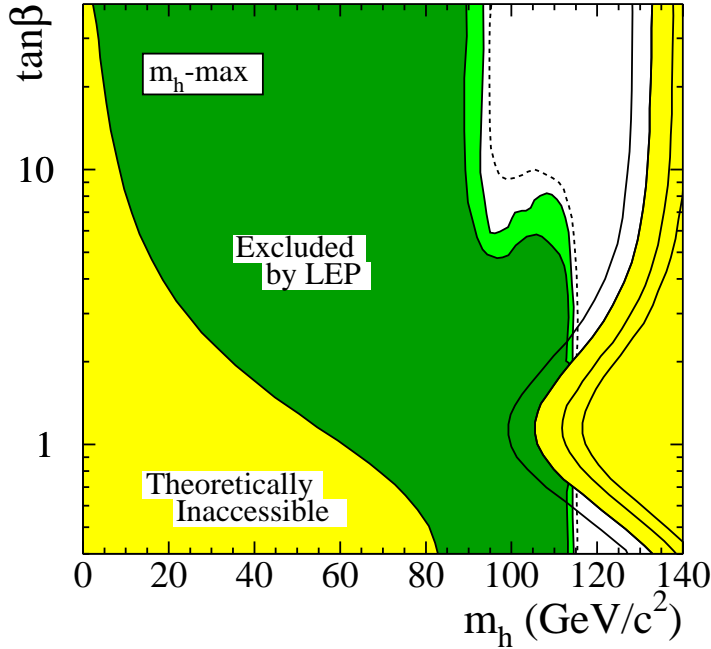,scale=1.1}
        \caption{\it MSSM exclusion contours at 95\% CL (light shading) and 97\% (dark shading) as a
                function of $\tan\beta$ and lightest Higgs boson mass $M_{h^0}$ \cite{exp_ADLO2006_MSSM}.}
        \label{fig:exp_MSSM}
\end{figure}

Like for the Standard Model Higgs boson search, Higgs searches in supersymmetric theories
were performed by the four LEP Collaborations, including all LEP2 data up to the highest 
energy of 209~GeV. The combined LEP data \cite{exp_ADLO2006_MSSM} show no significant signal for
Higgs boson production, neither in Higgs-strahlung processes $e^+e^- \to Zh^0/H^0$
nor in associated neutral or charged pair production $e^+e^- \to A^0 h^0/H^0, H^+ H^-$. 
These null results are used to set upper bounds on topological cross sections for a number of
Higgs-like final states. Furthermore, they are interpreted in
a set of representative MSSM benchmark models \cite{exp_benchmarks}, with
and without $CP$-violating effects in the Higgs sector. \\[-1mm]
 
Here, as a characteristic example the $\mit{m_{h}-max}$ scenario \cite{vhgg:susy-mh-max} 
of the MSSM has been chosen for illustration where the stop mixing parameter 
is set to a large value, $X_t = 2M_{susy}$. This model is designed to
maximize the theoretical upper bound on the $h^0$ mass within the MSSM for a given
mixing parameter $\tan\beta$ and fixed top and supersymmetry mass parameters, 
$M_t$ and $M_{susy}$. The model thus provides the largest parameter space 
in the $h^0$ direction and conservative exclusion limits for $\tan\beta$.
The exclusion contours from the LEP2 combination for the MSSM
are shown in Figure \ref{fig:exp_MSSM} (Ref.\cite{exp_ADLO2006_MSSM})
for the lightest Higgs boson mass $M_{h^0}$ and the mixing parameter $\tan\beta$. \\[-1mm]

The lower bound for the mass of the charged Higgs bosons $M_{H^\pm}$ has been set
at LEP to about 78.6~GeV, significantly below the beam energy as the cross section for
scalar pair production is suppressed near threshold. Similar bounds of 92.8 and 93.4~GeV
apply to the masses of the heavy $CP$-even and odd neutral Higgs bosons $M_{H^0,A^0}$,
respectively, depending on the mixing parameter, {\it cf.} \cite{PDG} for details. \\[-1mm]

\newpage

After accumulating high integrated luminosity at LHC, the MSSM Higgs sector can be 
observed in the production of the lightest Higgs boson $h^0$ and the heavy Higgs bosons 
$H^0, A^0, H^\pm$ up to 1 TeV. The discovery of the entire spectrum of five Higgs bosons
is possible only in part of the MSSM parameter space. \\[-1mm]

\vspace{4mm}

\section{R\'{e}sum\'{e}}

\noindent
Even though the Higgs bosons of the Standard Model or related extended theories have not been discovered 
at LEP, while the search continues fervently at LHC, experiments at the colliders could constrain the Higgs 
sector \cite{EB,Higgs123,HaGuKi}, as potentially realized in nature, quite strongly. \\[-1mm] 

Evaluating quantum corrections connected with high precision measurements at LEP, SLC in Stanford and 
the Tevatron collider, restricts the mass of the Higgs boson in the Standard Model \cite{elwWG} to values
\begin{eqnarray}
M_H &=&     92 ^{+34}_{-26} \, {\rm GeV}          \nonumber\\[2mm]
    &\leq& 161 \, {\rm GeV} \; (95\% \, {\rm CL})
\end{eqnarray}
Thus, small values are suggested for the mass if the Higgs boson is realized as a fundamental particle.
Likewise, this observation is compatible with values of the lightest Higgs-boson mass in supersymmetric 
extensions of the Standard Model. By contrast, technicolor in its simplest realization as a high-scale copy 
of QCD could be proven not compatible with the high-precision data. Complex constructs would be needed if 
electroweak symmetry breaking were triggered by novel strong interactions. \\

Direct searches for Higgs bosons in $Z$-decays at LEP1 and, primarily in the Higgs-strahlung process
$e^+e^- \to HZ$ at LEP2, have ruled out, on the other hand, the mass range \cite{exp_ADLO2003} from zero 
up to the 95\% CL bound of
\begin{equation}
M_H \geq 114.4 \ {\rm GeV}   
\end{equation} 
so that only a small gap has been left open by LEP. Correspondingly, large areas of parameter 
spaces in extended theories, as suggested by supersymmetry for instance, have been ruled out by the 
negative outcome of LEP searches. \\

The ATLAS and CMS collaborations at LHC have recently reduced the upper limit of the Higgs mass 
in the intermediate range of the Standard Model considerably \cite{publicCERNsem} compared 
with the limit derived from electroweak precision measurements. After excluding Higgs bosons 
in the high mass range up to 453/600 GeV at ATLAS/CMS, with a small cut-out in the first 
experiment, only a very narrow gap, {\it cf.} Fig.{\ref{fig:F_BlueBand}}, is left open, 
at the time of writing, for the Higgs mass in the intermediate range,
\begin{equation}
  115.5 / 115\; {\rm GeV}\;\; - \;\; 131 / 127\; {\rm GeV}\;\; {\rm at\;\; ATLAS / CMS}.
\end{equation} 
Intriguing hints have been observed by both experiments \cite{publicCERNsem} in the mass range 
of 124 - 126~GeV, particularly in the resonating $H \to \gamma\gamma$ channel. 
Though the excess in both experiments is presently not strong enough to claim a discovery, 
the coming year 2012 can reasonably be expected to lead us to the final decision. \\[-1mm]

This gap is close to the minimum of the $\chi^2$ distribution in Fig.~{\ref{fig:F_BlueBand}}
describing the predictions for the Higgs mass derived from high precision experiments 
-- a potential triumph of the Standard Model. \\[-1mm] 

With high luminosity at the LHC, the experiments have the potential to cover also the entire
large Higgs mass range of the Standard-Model. They also exhaust the minimal supersymmetry parameter 
space by observing the production of the light $h^0$, and $H^0, A^0, H^\pm$ up to 1 TeV, though 
the complete spectrum of five states may only be accessible in part of the parameter space. \\[-1mm]

Within the framework of the Standard Model LHC covers the entire mass range possible for the Higgs
boson. In this way, the closure 
of the Standard Model as a renormalizable theory including only a few basic parameters can be achieved 
or falsified. The outcome of the LHC experiments will decide whether electroweak symmetry is broken 
in a weakly coupled fundamental Higgs sector or, potentially, through spontaneous symmetry breaking 
by novel strong interactions. Each realization suggests extensions of the Standard Model either 
in a sector of fundamental fields, for which supersymmetry is the preferred paradigm, or in a sector 
of new strong interactions. In any case, perspectives of new structures in nature are opened 
at small distances by unraveling experimentally the mechanism of electroweak symmetry breaking. \\
 
\vspace{8mm}
\noindent
{\bf Acknowledgements} \\

\noindent
{\it Communications with J.~Erler, D.~Haidt, W. Kilian, M.~Kr\"amer, M.~Spira and D.~Zerwas
are gratefully acknowledged, likewise diagrams adopted from A.~Djouadi. Special thanks 
go to P.~Igo-Kemenes, former convener of the LEP Higgs Working Group, for valuable 
remarks on the manuscript, and to M.~Gr\"unewald for providing us with the new version 
of the Blue-Band-plot including the most recent LHC exclusion bounds for the Higgs mass 
in the Standard Model. We also thank the editor, Professor H.~Schopper, for the invitation
to this report and his continuous advice.}


\end{document}